# Naturally-meaningful and efficient descriptors: machine learning of material properties based on robust one-shot *ab initio* descriptors

Sherif Abdulkader Tawfik[1,2*] and Salvy P. Russo[1,3*]

[1]ARC Centre of Excellence in Exciton Science, School of Science, RMIT University, Melbourne, VIC 3001 Australia

[2]Institute for Frontier Materials, Deakin University, Geelong, Victoria 3216, Australia

[3]Chemical and Quantum Physics, School of Science, RMIT University, Melbourne VIC 3001, Australia

**Abstract**

Establishing a data-driven pipeline for the discovery of novel materials requires the engineering of material features that can be feasibly calculated and can be applied to predict a material's target properties. Here we propose a new class of descriptors for describing crystal structures, which we term Robust One-Shot *Ab initio* (ROSA) descriptors. ROSA is computationally cheap and is shown to accurately predict a range of material properties. These simple and intuitive class of descriptors are generated from the energetics of a material at a low level of theory using an incomplete *ab initio* calculation. We demonstrate how the incorporation of ROSA descriptors in ML-based property prediction leads to accurate predictions over a wide range of crystals, amorphized crystals, metal-organic frameworks and molecules. We believe that the low computational cost and ease of use of these descriptors will significantly improve ML-based predictions.

## 1. Introduction

A major objective in material science is to generate machine learning (ML) models that can accurately, and rapidly, predict a property for a given material by using information derived from the material's structure only.[1,2] Predicting material properties such as the energy bandgap would then only take a few seconds or a fraction of a second using an ML model, instead of consuming several hours, or even days on a supercomputer to perform a first principles calculation, such as density functional theory (DFT). With the availability of massive materials datasets such as MaterialsProject[3] and AFLOW[4] which host more than 3.5 million materials, it is becoming increasingly possible to screen materials for their properties.[2] To achieve such an objective, one must find features that can map a material structure against the highly nonlinear material properties. The vector of feature descriptors (the individual quantities that constitute the feature)[1] must be unique to each material and feasible to calculate. An ML model can subsequently be trained to translate descriptors into properties i.e. perform the mapping of structure against property. No matter how sophisticated or "deep" the ML models are, they will fail as long as the descriptors are poorly chosen.

---

[1] The terms "features" and "descriptors" are used interchangeable in the literature. Here we refer to a "feature" as a group of "descriptors". Note that other terms, including "attributes" and "fingerprints", are also frequently used in the literature, and they have the same meaning as "descriptors".

The quality of descriptors is usually appraised by the ability of the descriptors to train predictive ML models. However, we emphasize the importance of three other key elements for judging the quality of descriptors, that are as important as their predictive power:

1. Meaningfulness of features: the term "meaningful features" appears frequently in the broad ML literature, such in in Ref. [5]. In the context of material science, the term loosely means that the features are related to a physical and/or chemical principle. An example is Ref. [6].
2. Calculation efficiency: the cost of computing the feature should be much less than that of calculating the target property.
3. Number of descriptors within a feature: the expression of a material structure into a relatively small number of features (i.e. a few hundreds) can ensure the simplicity of the ML model. Features that require the calculation of thousands of features entail costly storage requirements for datasets, higher processing requirements for the utilization of the trained ML models, and non-transparent, or "black-box" ML models. [5]

We call these four criteria of ML features for materials the **MENA criteria** (Meaningful, Efficient, small Number of descriptors, Accurate). A number of ML features have recently been proposed in the literature for predicting the various DFT-calculated properties for materials, but they differ in how they satisfy the MENA criteria. We classify these features into the following four classes:

1. *Elemental features*: this is the simplest type of features, and the quickest to calculate. The descriptor values within these features are directly related to a property of the elements within the crystal structure or molecule, and therefore are physically and chemically meaningful. For example, for a crystal structure, a possible elemental feature is the mean atomic number and mean elemental melting point of the atoms within the crystal unit cell. However, these features are nonunique; two materials with equal composition, but different structural phases, will have the same elemental features. They are thus not accurate. It was also reported that that, in some cases, the most significant features for predicting a property seem to be counter-intuitive.[6] Using those features alone might work in limited cases, such as when using a small dataset (such as the ~300 materials dataset in Ref. [7]), but cannot be generalized for the broader set of materials. The reason these features work is related to the distribution of polymorphs in present-day materials databases: in MaterialsProject, for example, the average number of polymorphs for each materials is ~1.4. If there were more polymorphs, elemental features might suffer from the fact that a material will have multiple property values (such as SiC, which has 27 polymorphs in MaterialsProject and their bandgaps range from 0 eV to 2.3 eV).
2. *Geometry-based features*: these include property-labelled materials fragments (PLMFs) [8], crystal graphs[9] and symmetry functions,[10] among others. These features calculate translationally-invariant geometric, as well as elemental, quantities based purely on the material's geometry. A simple, yet effective geometric descriptor is the symmetry group of the material's lattice, which is ideally hot-coded into 230 separate zero-or-one columns. Many of the features in this class are mathematically very complex. An example of such descriptors is the symmetry functions. They are evaluated from the summation of exponential functions of the atomic distances. The features in this class are generally feasible to calculate.
3. *Electronic structure features*: they are calculated based on the electronic properties of the individual constituents of the material or molecule. That is, they are derived for separate atoms and/or bonds within the structure, but not for the entire structure. Examples include: the electronic structure attributes,[11] molecular orbital attributes[12] and smooth overlap of atomic positions (SOAP).[13]
4. *Ab initio*-based features: examples are the molecular orbital energies[14] and the OrbNet Denali descriptors.[15] These descriptors are calculated by performing a full *ab initio* calculation at a low level of theory, and then the output of this calculation is used as descriptors for predicting the system's properties at a higher level of theory; that is, ML here is "correcting" the outcome of low-level theory. A related ML procedure is the Δ-ML,[16,17] in which an ML model is trained to predict the difference in the value of a property, such as the HOMO-LUMO gap, between the value obtained using a high level of theory, and that using a low level of theory. Using ML to correct the results of DFT calculations has been known for a while.[18] The descriptors in these features are highly meaningful, since that they correspond directly to physically-computed quantities. In fact, the meaning of the descriptor values overlap with that of the target properties, such as the case of using molecular orbital energies (HOMO and LUMO values) to train a model to predict the HOMO-LUMO gap. However, this class of features is the most expensive to calculate; a DFT calculation scales as $N^3$, where $N$ = number of atoms in the unit cell.[19] Thus, the calculation of the descriptors in these features typically require the utilization of high-performance computing facilities and long computing hours.

In this work we propose a new feature that can achieve the MENA criteria. Inspired by the *ab initio*-based features class, our *robust one-shot ab initio* (ROSA) descriptors of a given material are DFT-based descriptors. However, unlike the current member features of this class, ROSA descriptors are not calculated self-consistently; in fact, they are calculated by performing only one step in the self-consistent field (SCF) iteration. The ROSA descriptor values include the eigenvalues and total energy components that result from this computational step. ROSA descriptors are more computationally efficient than other *ab initio*-based features, equally meaningful, can be expressed with a small number of descriptors (109 descriptors, as will be explained below) and are highly accurate, as will be demonstrated in the following sections. We augment the ROSA descriptors with other material features, including atom-based and geometry features, and compare the predictive power of the different feature classes. We demonstrate the accuracy of predicting a range of material properties using the ROSA descriptors. These descriptors are also shown to be predictive for properties that are not directly related to a material's energy, such as the material's vibration properties. By using the energy bandgap, calculated using the popular Perdew-Burke-Ernzerhof (PBE) functional,[20] as an additional input quantity, we also demonstrate the accuracy of predicting high-level energy properties of materials, namely: the HSE bandgap,[21] GW bandgap[22] and the exciton binding energy calculated by solving the Bethe-Salpeter equation (BSE).[23] We display a schematic diagram of the ROSA descriptors in Figure 1(a).

The manuscript is organized as follows: Section 2 introduces the ROSA feature as well as the other features that will be used in this work, Section 3 presents the results of training machine learning models on a range of material properties: section 3.1 is for energetic, mechanical and vibrational properties of bulk material properties, section 3.2 is for the higher-level energetics of two-dimensional materials, and section 3.3 is for the prediction of properties molecular systems. Section 4 is the conclusion of the work.

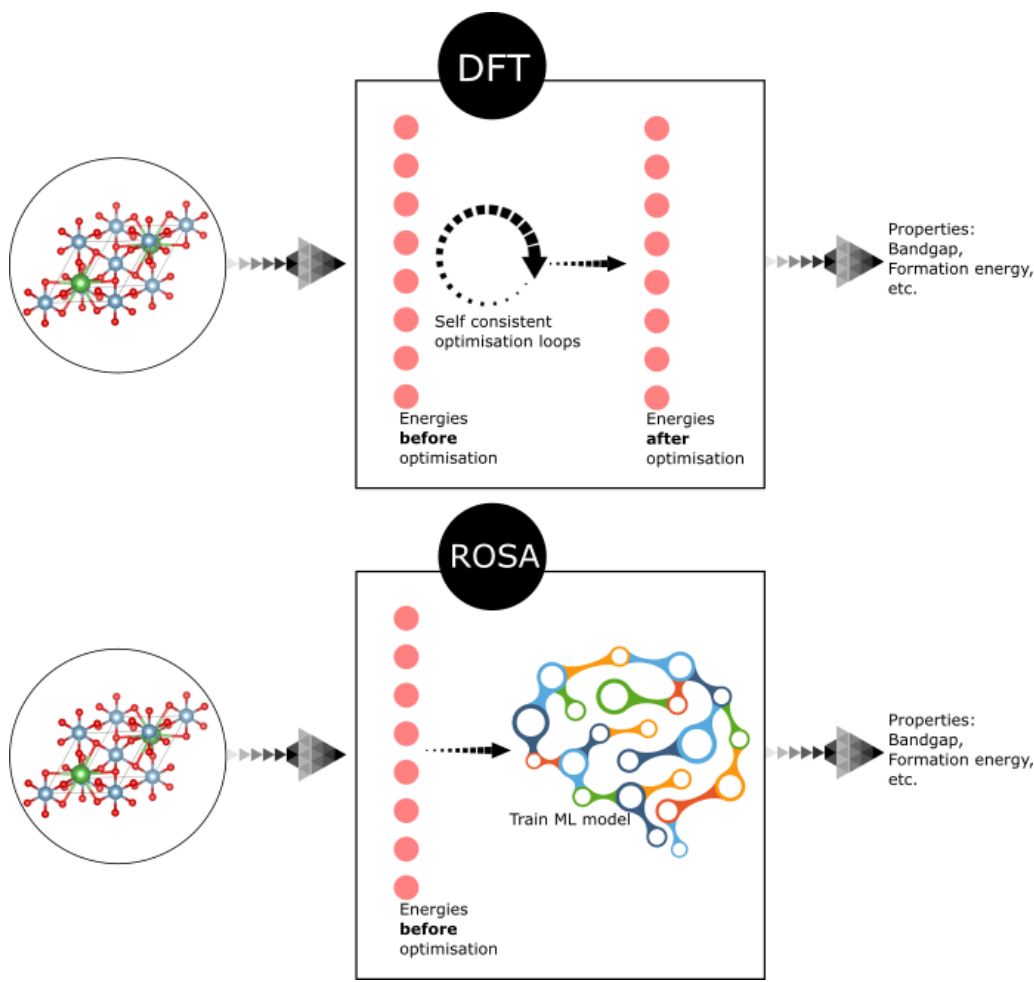

Figure 1: A schematic illustration of the ROSA feature. The top figure displays an ordinary DFT calculation, in which two self-consistent optimization loops (the electron density optimization and geometry optimization) determine the electronic structure of the ground state system. This ground state is identified by energy eigenvalues and total energy (the system "Energies"). The lower figure shows how ROSA feature approximate the properties of the ground state system: it extracts the system "Energies" before optimization and uses that to train a machine learning model to predict the desired properties.

We examine the calculation CPU times and the calculated ROSA eigenvalues, based on single-core calculations performed using Intel Xeon Scalable processor cores. For a sample of ~230 materials from the MaterialsProject with varying sizes, we compare the CPU time and the eigenvalues obtained by running a full SCF using VASP on a $4 \times 4 \times 4$ **k**-points mesh, the ROSA feature calculation using the linear combination of atomic orbitals (LCAO) (labelled ROSA LCAO) and the plane wave basis set (labelled ROSA PW), and the ROSA calculation using VASP version 5.4.4.[24] All of these calculations have been performed on a single core, to enable direct comparison of CPU times. The results are displayed in Figure 2(a). In the figure, while the CPU time of ROSA PW and ROSA LCAO nearly coincide for a large number of materials, ROSA PW takes more CPU time than ROSA LCAO due to the increased complexity of PW calculations in larger unit cells. For this reason, the LCAO basis sets are used in the present ROSA calculations. Moreover, the full SCF calculation consumes much more time than both ROSA PW and ROSA LCAO by 1 to 4 orders of magnitude, which shows the significant time saving that is achieved with ROSA descriptors.

In Figure 2(b), we compare the eigenvalues obtained using the different methods by calculating the absolute value of the difference between the 100 eigenvalues obtained using ROSA LCAO, and those obtained using ROSA PW, VASP PW and VASP full SCF. The eigenvalues of ROSA PW are close to those of ROSA LCAO, while the eigenvalues obtained using VASP full SCF and VASP PW are quite different from those obtained using ROSA LCAO.

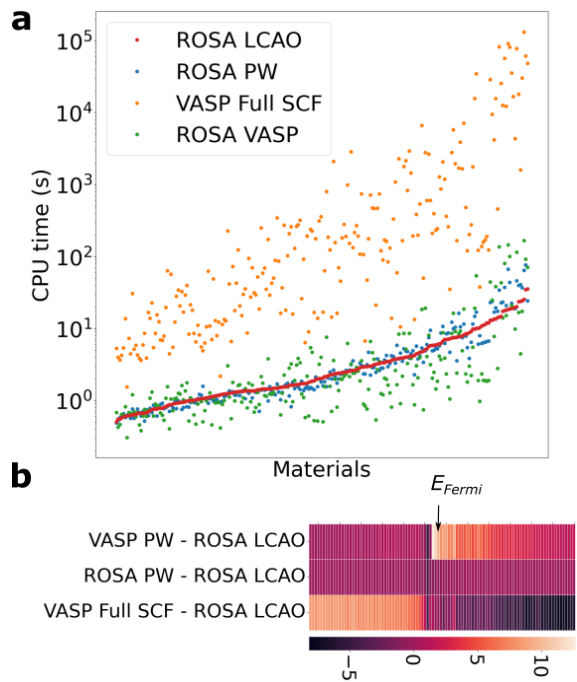

Figure 2: (a) A comparison between the ROSA descriptors calculated using LCAO the basis set, ROSA using the PW basis set, a full SCF calculation using PW basis set and the ROSA using VASP. (b) The absolute value of the difference between the eigenvalues of the ROSA descriptors and those calculated using VASP, ROSA PW and the full SCF. The values are in eV, and the position of the Fermi energy ($E_{Fermi}$) is indicated with an arrow.

## 2. Features

In this work we have utilized three groups of features: the ROSA feature, the atom-based statistical properties, and a modification of the symmetry functions introduced in Ref [10].

**2.1 ROSA feature**

Although the calculation of the descriptors in the ROSA feature requires a DFT calculation, it is a rather "superficial" calculation, in which the applied DFT is set at the lowest possible level of accuracy, and only one non-self-consistent SCF step is executed; that is, not a full SCF iteration. The descriptors obtained from this calculation are the energy eigenvalues and total energies. We use the python-based GPAW code (version 21.1.0),[25] which can be easily installed on personal computers, for these calculations. The spin-restricted PBE exchange-correlation function is used, and calculations are performed at the Γ point. Note that setting the maximum iteration to 1 without raising an error from the present version of GPAW required that the source code be modified. This ROSA calculation produces a unique set of eigenvalues and 9 total energy values for each material. We centre each set of eigenvalues at the Fermi level and consider the eigenvalues corresponding to 50 occupied and 50 unoccupied orbitals.

The present implementation of the ROSA feature relies on pseudopotentials currently available in the GPAW library, which does not include lanthanides and actinides. If the pseudopotential of any atom in the unit cell is not available, the algorithm substitutes it with that of the yttrium atom. The choice of the yttrium atom is arbitrary, but it had a limited impact on the accuracy of the predictions of target properties in materials where a lanthanide/actinide was substituted with Y.

## 2.2 Basic atom and crystal descriptors (BACD)

This set of descriptors includes the properties of the individual atoms in the crystal, as well as the symmetry information of the crystal structure. The atom-based descriptors include 21 elemental descriptors such as the bulk modulus, ionic radius, rigidity modulus and the molar volume. For each of these descriptors, we include four statistical values over all atoms within the crystal: mean, standard deviation, maximum and minimum values. The crystal structure descriptors include the symmetry group, which is hot-encoded into 230 columns to denote each of the 230 symmetry groups, and the statistical summary of the distance matrix. The distance matrix is the set of distances between each atom and its neighbouring atoms within the crystal. In the present implementation there are 325 BACD features.

## 2.3 Symmetry functions (G)

We calculate the following two symmetry functions for the crystal:

$$G_1^i = \sum_j e^{-\rho(R_{ij}-R_s)^2} f_c(R_{ij})$$

$$G_2^i = \sum_j e^{-\rho(R_{ij}-R_s)^2} e^{-\gamma|Z_i-Z_j|} f_c(R_{ij})$$

$$f_c(x) = \begin{cases} 1, & x < R_c \\ 0, & x \geq R_c \end{cases}$$

where $R_{ij}$ is the distance between the atoms at positions $R_i$ and $R_j$. The $G_1^i$ and $G_2^i$ symmetry functions are calculated for each atom $i$ in the crystal, $G_1^i$ is inspired by the function with the same label in Ref. [10], whereas $G_2^i$ is a modification of $G_1^i$ in which a term containing the difference in atomic numbers, $-\gamma|Z_i - Z_j|$, is added to the exponential function. The cutoff function $f_c(R_{ij})$ takes a simpler form than that in Ref. [10]. The variables $\rho$, $R_s$, $\gamma$ and $R_c$ are descriptor parameters. For each crystal structure, the mean value of $G_1^i$ for all atoms $i$ in the crystal represents one descriptor in the feature set. This value is obtained for a specific choice of the variables $\rho$, $R_s$, and $R_c$. Thus, the full descriptor set for the $G_1^i$ function are obtained by assigning different values of $\rho$, $R_s$, and $R_c$ and calculating $\sum G_1^i / N$, where N is the number of atoms in the crystal. The same procedure is applied for the $G_2^i$ function, which has the additional parameter $\gamma$. In total we generate 600 G descriptors for the subsequent machine learning procedures. Thus the total number of descriptors, including all three classes, is 1,034.

Each of the three descriptor classes is further divided into groups to simplify the analysis. The ROSA descriptors are divided into 3 groups: VBM (the occupied energy levels), CBM (the unoccupied energy levels), and *e* (the total and constituent energy components including: the total exchange-correlation energy, total kinetic energy, total Hartree energy, Fermi energy, total Coulomb energy, total entropy, total electron-atom interaction energy and total free energy). The BACD descriptors are divided into 5 groups: SG (symmetry group), T (thermal properties of the elements), Geo (geometric features including the distance matrix and the lattice angles) and A (other atomic features). The G descriptors are divided into 2 groups: G1 (the $G_1$ features) and G2 (the $G_2$ features). The analysis of the feature importance will involve the aggregation of the individual descriptor groups, as is shown in Figure 3(a).

# 3. Results and discussion

### 3.1 Energetics, mechanical and vibrational properties of bulk systems

The ROSA feature includes approximate information about the PBE bandgap and total energy of the system, and therefore, supported by the BACD and G features, an ML model should be able to accurately map them to the converged PBE and formation energy of the system. We construct a dataset of 65,899 materials from MaterialsProject, of which 24,311 are semiconductors (37%). We consider semiconductors as materials with a bandgap > 0.1 eV, and metals are otherwise. We train XGBOOST regression and classification models, and for all models we use 80% of the dataset as a training set, and reserve the remaining 20% as a test set.

First we apply the three descriptor classes for the prediction of three energetic properties: the PBE bandgap, formation energy per atom, $E_f$ and the bulk modulus, $K_{VHR}$. For the PBE bandgap prediction, we train a classifier model to distinguish metallic from semiconducting crystals, and a regression model to predict the bandgap values for the materials. The classifier achieves an area under the curve (AUC) of 0.97. The regression model achieves an $R^2 = 0.89$ with a mean absolute error (MAE) of 0.22 eV (Table 1). We train another regression model for the prediction of the formation energy per atom, and it achieves a very high accuracy: $R^2 = 0.97$ with a mean absolute error (MAE) of 0.11 eV. These two accuracy values compare well against those reported in Ref. [8]. The receiver operating characteristic (ROC) of the classifier model is displayed in Figure 3(b), and the correlation plots for the regression models are displayed in Figure 3(c-d).

The $K_{VHR}$ dataset currently includes 13,147 values in MaterialsProject, and the dataset for the vibrational properties includes 1,521 materials in MaterialsProject (which goes down to 1,245 materials after removing structures with imaginary phonon frequencies). Upon training an XGBOOST regression model on predicting $K_{VHR}$ values, the model achieves an $R^2 = 0.86$, MAE = 16 GPa (Table 1). While the accuracy of this model is not as high as the model trained on the PLMF features in Ref. [8], which achieved $R^2 = 0.97$, the accuracy reported here is still a significant one because it covered a larger set of data than the dataset in Ref. [8].

In the feature importance matrix displayed in Figure 3(a), the feature importance is calculated from the *total gain* in the XGBOOST model tree. The key observation in the matrix is that the ROSA descriptors are the most significant contributor to the model prediction of the PBE bandgap $E_G$, and the third most significant contributor for $E_f$. As would be expected, the atomic features (A) are the prime contributor for $E_f$, followed by geometric features of the crystal. The primary ROSA descriptor group that contributes to the prediction of $E_f$ is the set of total energy values in the pristine system, $e$, where as for the prediction of $E_G$ are the sets of VBM and CBM descriptors.

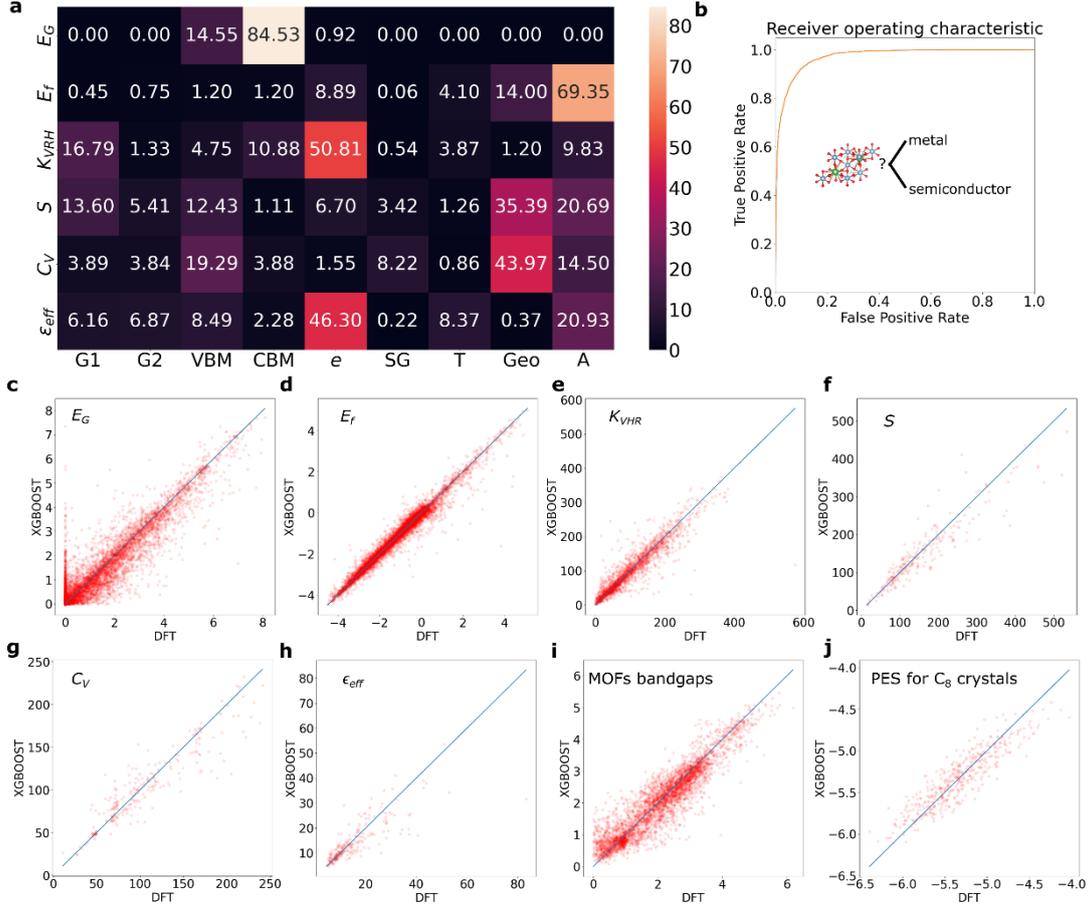

Figure 3: Performance of the three feature classes (ROSA, BACD and G) in the prediction of the bandgap, $E_G$, formation energy by atom, $E_f$, bulk modulus, $K_{VRH}$, the vibrational entropy, $S$, the specific heat, $C_V$ and the effective dielectric constant, $\varepsilon_{\text{eff}}$. (a) The feature importance matrix for predicting the properties by the feature groups outlined in Section 2. (b) The receiver operating characteristic (ROC) for the metal/insulator classification model (described in the text). (c-j) The correlation plots for the prediction of the regression models for $E_G$, $E_f$, $K_{VRH}$, $S$, $C_V$, $\varepsilon_{\text{eff}}$, the bandgaps for the materials in the QOMF database and the potential energy surfaces (PESs) for amorphized diamond unit cells, respectively.

In order to examine the generality of the ROSA descriptors to non-energetic properties, we train regression models to predict three quantities that are obtained from the vibrational spectrum of a material: entropy, $S$; specific heat, $C_V$; and the effective polycrystalline dielectric function, $\varepsilon_{\text{eff}}$ defined as

$$\varepsilon_{\text{eff}} = \frac{3\varepsilon_x \varepsilon_y \varepsilon_z}{\varepsilon_x \varepsilon_y + \varepsilon_x \varepsilon_z + \varepsilon_y \varepsilon_z},$$

where $\varepsilon_i$ is the $i$ component of the dielectric tensor, $i=x,y,z$. The achieved accuracy of the trained regression models are $R^2 = 0.85$, MAE = 23 meV/atom/K; $R^2 = 0.85$, MAE = 13 meV/atom/K and $R^2 = 0.66$, MAE = 3.2, respectively. These values are comparable to those reported in Ref. [26].

Table 1: The $R^2$ and mean absolute error (MAE) values for the prediction of the quantities $E_G$, $E_f$, $K_{VRH}$, $S$, $C_V$, $\varepsilon_{\text{eff}}$, MOF bandgap for the QMOF structures and the total energy for amorphized $C_8$ crystals (potential energy surface, PES). The MAE values for $E_G$, $E_f$ and MOF bandgaps are in eV, for $K_{VRH}$ it is in GPa, for $S$ and $C_V$ in meV/atom/K and for the PES for $C_8$ crystals is eV/atom.

| Property | Dataset size | $R^2$ | MAE |
| --- | --- | --- | --- |
| $E_G$ | 65,899 | 0.89 | 0.22 |
| $E_f$ | 65,899 | 0.97 | 0.11 |
| $K_{VHR}$ | 13,147 | 0.86 | 16 |
| $S$ | 1,521 | 0.85 | 23 |
| $C_V$ | 1,521 | 0.85 | 13 |
| $\varepsilon_{eff}$ | 1,521 | 0.66 | 3.2 |
| MOF bandgap | 20,425 | 0.86 | 0.31 |
| PES for $C_8$ crystals | 5,000 | 0.81 | 0.14 |

The ROSA descriptors group $e$ plays the key role in determining $K_{VHR}$ values, as shown in Figure 2(a); it constitutes the most significant set of descriptors (sum of feature importance of features in the group is ~ 51%). Owing to the mechanical nature of the bulk modulus, the G1 feature group is the second most significant determinant of the property. The ROSA eigenvalue feature groups VBM and CBM together form the fourth most significant group. These results show the importance of ROSA descriptors for properties that are not directly related to the material's energy eigenvalues or energy components. Those descriptors are even significant in predicting the three vibrational properties $S$, $C_V$ and $\varepsilon_{eff}$. Given that these properties are strongly related to the material's structure and composition, the Geo, A, G1 and G2 feature groups are the two most significant determinants of all three properties as shown in Figure 3(a). The ROSA eigenvalue group are less significant in determining the three vibrational properties.

We also examine the ability of the ROSA descriptors in capturing the bandgaps in a different class of systems: metal-organic frameworks (MOFs) and amorphized crystals. We trained an XGBOOST model on the entire QMOF database,[27] which has 20,425 MOFs along with their PBE-calculated bandgaps. Here we used the ROSA and the BACD descriptors, and achieved a reasonable prediction accuracy, with $R^2 = 0.86$. This value for $R^2$ is close to that reported by Rosen et al.[27], where the highest achieved accuracy was $R^2 = 0.87$ which was obtained using the crystal graph convolutional neural network. For amorphized systems, we trained an XGBOOST model to predict the total energy of an amorphized diamond unit cells (with 8 atoms). The amorphization was performed by running a molecular dynamics in GPAW using the effective medium potential. A dataset of 5,000 snapshots of the carbon system were captured, and a single-point calculation for each of these structures was performed using VASP at the PBE level of theory. The trained model could predict the total energy/atom with an accuracy of $R^2 = 0.81$. These results assert the applicability of ROSA descriptors across different systems, quantities and calculation methods.

**3.2 High-level energetics of two-dimensional materials**

The largest open-source dataset that provides these bandgaps is C2DB,[28] which hosts > 4,000 2D materials and provides the HSE results for 1,302, GW bandgaps for 357 materials and exciton binding energies for 373 materials. Using this dataset as a training set, Liang and Zhu have recently trained ML models on ~150 2D materials to predict the HSE, GW and BSE energies of ~30 materials in the NREL dataset (which is composed of 3D materials) using a set of descriptors inspired by Phillips's ionicity theory,[29] claiming that they demonstrated transferable learning of these bandgap values. The prediction accuracy they reported was quite high when the PBE bandgap is added as a feature.

However, a line of best fit can be obtained for both GW and HSE bandgaps based on PBE bandgaps with $R^2$ of 0.95 and 0.88, respectively (see Figure 3(a,b)). The PBE bandgap therefore plays the role of the ROSA descriptor; it is a bandgap computed with a far less level of theory than the GW and HSE bandgaps. However, with such a strong linear dependence, the PBE bandgap value alongside other features will be the most dominant feature, as can be seen in the SHapley Additive exPlanations (SHAP) plots in Ref.[29].

We utilize the three descriptor classes to train regression XGBOOST models to predict the three quantities: HSE and GW bandgaps and the exciton binding energies, $E_b$. Note that the ROSA features that are used for predicting HSE and GW bandgaps and $E_b$ are the same as those that were used in Section 3.1, except that the PBE bandgap is added to the former. For the HSE and GW bandgaps, the prediction accuracy is enhanced with respect to the linear fits for the HSE and GW bandgaps: the regression models achieve $R^2$ = 0.95, MAE = 0.24 eV and $R^2$ = 0.98, MAE = 0.24 eV, respectively (Table 2). The significance of the descriptor groups is displayed in Figure 3(c). As expected, the DFT bandgap is a very significant feature in determining the HSE and GW bandgaps, where the feature importance > 90%.

Table 2: The $R^2$ and mean absolute error (MAE) values for the prediction of the PBE bandgap ($E_{PBE}$), the HSE bandgap with ($E_{HSE}^{PBE}$) and without ($E_{HSE}$) using the PBE bandgap as a feature, the GW bandgap with ($E_{GW}^{PBE}$) and without ($E_{GW}$) using the PBE bandgap as a feature, and the exciton binding energy $E_b$. The MAE values are in eV.

| Property | Dataset size | $R^2$ | MAE |
|---|---|---|---|
| $E_{PBE}$ | 1,302 | 0.90 | 0.16 |
| $E_{HSE}^{PBE}$ | 1,302 | 0.95 | 0.23 |
| $E_{HSE}$ | 1,302 | 0.86 | 0.37 |
| $E_{GW}^{PBE}$ | 357 | 0.98 | 0.26 |
| $E_{GW}$ | 357 | 0.93 | 0.37 |
| $E_b$ | 373 | 0.89 | 0.18 |

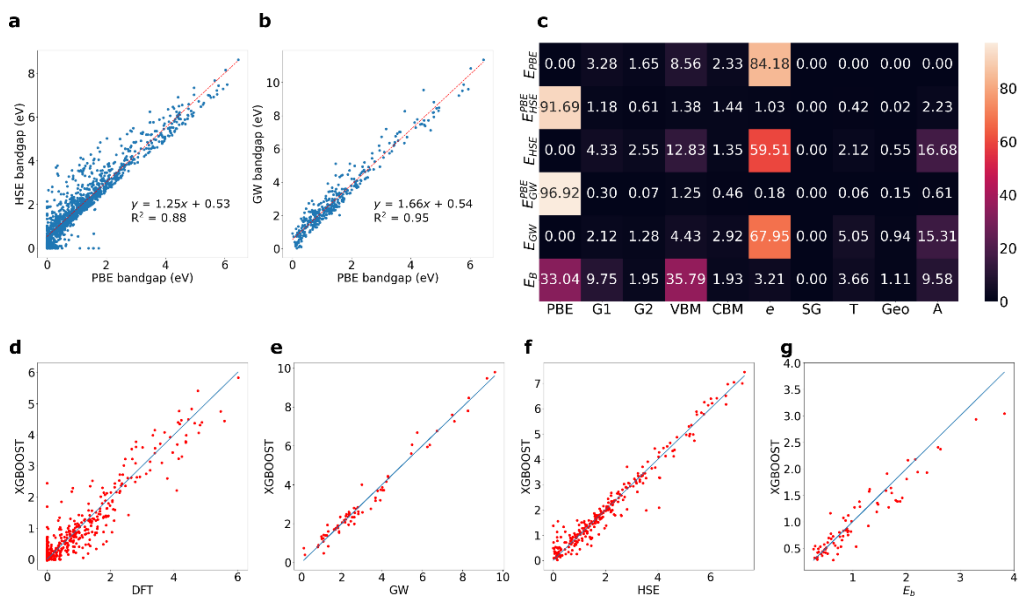

Figure 4: (a) The feature importance matrix for predicting the properties by the descriptor groups outlined in Section 2. The properties shown include the PBE bandgap, $E_{PBE}$, the HSE bandgap with ($E_{HSE}^{PBE}$) and without ($E_{HSE}$) using the PBE bandgap as a feature, and the GW bandgap with ($E_{GW}^{PBE}$) and without ($E_{GW}$) using the PBE bandgap as a feature, and the exciton binding energy $E_b$. (b,c) The correlation plots for and linear fitting for the HSE and GW bandgaps versus the PBE bandgaps. (d-h) The correlation plots for the prediction of the regression models for $E_{PBE}$, $E_{HSE}$, $E_{GW}$ and $E_b$, respectively.

When the PBE bandgap is removed as a descriptor from the training sets, the ROSA descriptors become the most significant features for predicting the HSE and GW bandgaps. As can be seen in Figure 3(c), the *e* descriptor group has the highest significance, followed by the A group. These regression models achieve $R^2$ = 0.86, MAE = 0.37 eV and $R^2$ = 0.93, MAE = 0.37 eV, respectively. The *e* descriptor group also has the highest significance when we train a model to predict the PBE bandgap values, followed by the A group. The VBM group comes at the third place in terms of significance in predicting the $E_{PBE}$, $E_{HSE}$ bandgaps and the fourth most significant in predicting the $E_{GW}$ bandgaps.

The $E_b$ does not directly correlate with the PBE bandgap, and therefore the bandgap is not expected to dominate the rest of the descriptors during ML training. The achieved accuracy for predicting $E_b$ is $R^2$ = 0.90 and MAE = 0.18 eV (Figure 3(g)). According to the aggregated feature importance heatmap in Figure 3(a), the PBE bandgap is the most significant features, but with a significance that is ~34%, unlike the >90% significance in the case of the HSE and GW bandgaps. The *VBM* replaces *e* as the most significant set of ROSA descriptors, and the G1 and G2 descriptor groups become more pronounced than in the prediction of any of the HSE/GW bandgaps. This shows the importance of adding ROSA descriptors of varying degrees of accuracy to improve the prediction accuracy of a complex energy quantity, $E_b$. By adding the bandgap of the optimized DFT calculation at the PBE level, as well as the non-self-consistent eigenvalues and energies (ROSA), the aforementioned accuracy was achievable.

### 3.3 Prediction of properties of molecular systems

As the forgoing analysis has been limited to crystal materials, we examine the applicability of the ROSA descriptors to molecular systems. We trained a XGBOOST model on the entire 134k stable small organic molecules, [30] for the prediction of three molecular quantities: the HOMO-LUMO gap $E_{HOMO-LUMO}$ (in eV), the isotropic polarizability α (in Bohr$^3$) and the free energy $G$ (in eV). The calculations of these properties were performed at the hybrid exchange DFT/B3LYP level of theory. We build the dataset using only the ROSA descriptors VBM, CBM and *e*. The XGBOOST model that was trained on 80% of the data achieved a very high prediction accuracy for the three descriptors, as displayed in Table 3: the prediction of $E_{HOMO-LUMO}$ achieved $R^2$ = 0.97, MAE = 0.15 eV; α achieved $R^2$ = 0.97, MAE = 0.15 Bohr$^3$; $G$ achieved $R^2$ = 0.97, MAE = 0.15 eV. The feature importance and correlation plots are displayed in Figure 5. The CBM descriptors are the key determinants of $E_{HOMO-LUMO}$ as is the case of $E_G$ in Figure 3(a), while *e* descriptors are the key determinants $G$, as is the case of $E_f$ in Figure 3(a), as well as α. Thus, ROSA descriptors alone are superior predictors of the electronic properties of molecular systems.

Table 3: The $R^2$ and mean absolute error (MAE) values for the prediction of the HOMO-LUMO gap $E_{HOMO-LUMO}$ (in eV), the isotropic polarizability α (in Bohr$^3$) and the free energy $G$ (in eV).

| Property | Dataset size | $R^2$ | MAE |
|---|---|---|---|
| $E_{HOMO-LUMO}$ | 133,706 | 0.97 | 0.15 |
| α | 133,706 | 0.98 | 0.85 |
| $G$ | 133,706 | 1.00 | 11 |

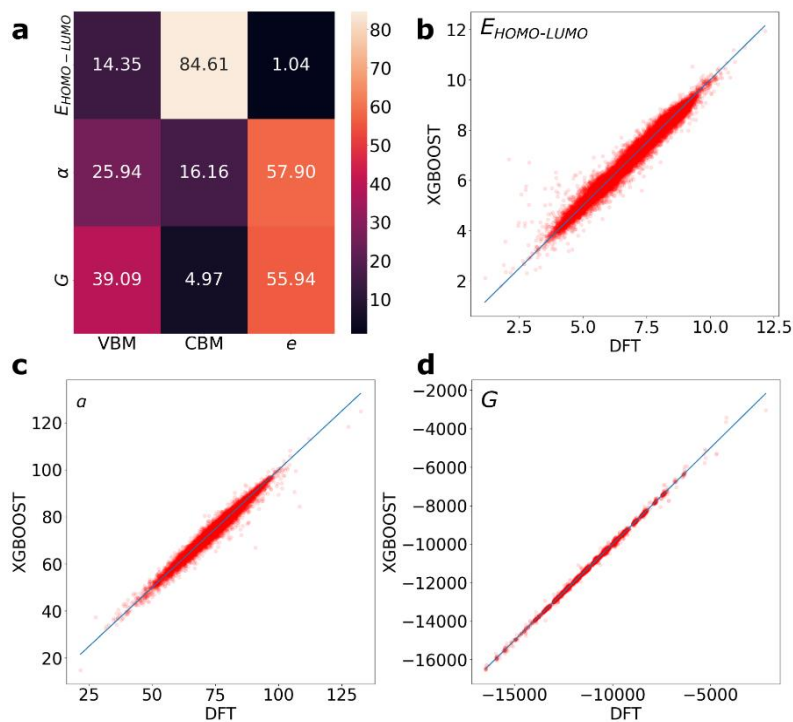

Figure 5: (a) The feature importance matrix for predicting the properties by the descriptor groups outlined in Section 2. The properties shown include the (b) HOMO-LUMO gap $E_{HOMO-LUMO}$ (in eV), (c) the isotropic polarizability α (in Bohr$^3$) and (d) the free energy $G$ (in eV).

# 4. Conclusion

We introduced a novel approach to generate naturally meaningful and computationally efficient features for crystal structures with a small number of descriptors and high predictive power. The robust one-shot *ab initio* (ROSA) features are calculated by performing only one step of the self-consistent field (SCF) calculation within a density functional theory iteration, and obtaining the eigenvalues and total energies as descriptor values. We demonstrated that ROSA descriptors can accurately predict key material quantities such as the energy bandgap and the bulk modulus. They were particularly a significant factor in accurately predicting the energy bandgap of metal-organic frameworks, the HOMO-LUMO gap of organic molecules and the potential energy surface of amorphized carbon crystals. An ML model trained on ROSA descriptors to predict a bandgap is mainly playing the role of the original, yet computationally expensive, SCF algorithm, rather than a deep learning black box. We also showed that the ROSA features, combined with atom-based and geometry features, are useful for the prediction of thermal properties.

# Abbreviations

| DFT | Density functional theory |
|-----|---------------------------|
| ML | Machine learning |
| MENA | Meaningful, Efficient, small Number of descriptors, Accurate |
| PLMF | property-labelled materials fragments |
| HOMO | Highest occupied molecular orbitals |
| LUMO | Lowest unoccupied molecular orbitals |

| SCF | self-consistent field |
|---|---|
| ROSA | Robust one-shot *ab initio* |
| HSE | Heyd–Scuseria–Ernzerhof |
| GW | An approximation method for the self-energy of many-body systems, where G stands for Green's function, W Coulomb interaction |
| BSE | Bethe-Salpeter equation |
| LCAO | linear combination of atomic orbitals basis set |
| VBM | Valence band maximum |
| CBM | Conduction band minimum |
| PW | Plane wave basis set |
| VASP | Vienna Ab initio Simulation Package |

# Competing Interests

The Authors declare no Competing Financial or Non-Financial Interests.

# Code availability

The python code that was used for generating BACD, ROSA and geometry descriptors is available in https://github.com/sheriftawfikabbas/crystalfeatures.

# Data Availability

The XGBOOST machine learning models as well as the scaled test sets are available in https://github.com/sheriftawfikabbas/crystalfeatures/tree/master/data.

# Author Contribution

S.A.T performed the calculations and data analysis. Both authors discussed the results and wrote the manuscript.

# Acknowledgements

S. A. T. recognizes the support of the Alfred Deakin Postdoctoral Research Fellowship from Deakin University. This work was supported by the Australian Government through the Australian Research Council (ARC) under the Centre of Excellence scheme (project number CE170100026). This work was also supported by computational resources provided by the Australian Government through the National Computational Infrastructure National Facility and the Pawsey Supercomputer Centre. This research used resources of the National Energy Research Scientific Computing Center (NERSC), a U.S. Department of Energy Office of Science User Facility located at Lawrence Berkeley National Laboratory, operated under Contract No. DE-AC02-05CH11231.

# Corresponding author

Correspondence to Sherif Abdulkader Tawfik: **s.abbas@deakin.edu.au**.

# Declarations

**Ethics approval and consent to participate**

Not applicable.

**Consent for publication**

The authors consent to the publication of this work.

**Funding**

There is no funding for this work.